\documentclass{appolb}%
\usepackage{epsfig}
\usepackage{rotating}
\usepackage{hyperref}
\usepackage{comment,color}
\usepackage{braket}
\usepackage{amssymb,amsmath,bm}
\usepackage{setspace,lipsum}
\usepackage{amsmath}
\usepackage{amsfonts}
\usepackage{amssymb}
\usepackage{graphicx}%
\begin{document}

\title{Heavy glueballs: status and large-$N_{c}$ width's estimate}
\author{F.~Giacosa$^{a,b}$
\address{$^a$Institute of Physics, Jan Kochanowski University, 25-406 Kielce, Poland}
\address{$^b$Institut f\"ur Theoretische Physik, Johann Wolfgang Goethe-
Universit\"at, 60438 Frankfurt am Main, Germany} }
\maketitle

\begin{abstract}
Glueballs, an old and firm prediction of various QCD approaches (lattice QCD,
bag models, AdS/QCD, effective models, etc.), have not yet been experimentally
confirmed. While for glueballs below $2.6$ GeV some candidates exist, the
situation for heavy glueballs (above $2.6$ GeV) is cloudy. Here, after a brief
review of scalar, tensor, and pseudoscalar glueballs, we present predictions
for the decays of a putative pseudotensor glueball with a lattice predicted
mass of $3.04$ GeV and a putative vector glueball with a lattice predicted
mass of $3.81$ GeV. Moreover, we discuss in general the width of heavy
glueballs by using large-$N_{c}$ arguments: we obtain a rough estimate
according to which the width of a glueball (such as the vector one) is about
$10$ MeV. Such a width would be narrow enough to enable measurement at the
future PANDA experiment.

\end{abstract}


\section{Introduction}

Gluons, the force carriers of strong interaction between quarks, carry
themselves color charge, hence they are expected to form white bound state,
called glueballs. The search for glueballs in the mesonic spectrum of the PDG
\cite{pdg}, is a long-standing activity, see the reviews in Refs. \cite{revgl}.

Theoretically, bag models \cite{bag} were the first to predict a spectrum of
glueballs. Later on, various other approaches have followed, e.g. QCD sum
rules, flux-tube model, Hamiltonian QCD, anti De-Sitter/QCD methods
\cite{variemass}. A reliable approach is lattice QCD
\cite{mainlattice,gregory} (both quenched and unquenched), in which a spectrum
of glueballs has been evaluated by a numerical simulation of QCD, see Table 1.

In conclusion, there is nowadays a (theoretical) consensus about the existence
of glueballs and the qualitative form of the spectrum, but up to now no
resonance could be \textit{unambiguously} classified as a predominantly
gluonic. In these proceedings, after a brief review of some glueball's
candidates, we present some recent developments for the decays of the heavy
vector and pseudotensor glueballs. Moreover, we present a new,
\textquotedblleft heuristic\textquotedblright, estimate of the glueball's
width by using large-$N_{c}$ arguments.

\begin{center}
\textbf{Table 1}: Central values of glueball masses from lattice QCD
\cite{mainlattice}.%

\begin{tabular}
[c]{|c|c|c|c|c|}\hline
$J^{PC}$ & Value [GeV] &  & $J^{PC}$ & Value [GeV]\\\hline
$0^{++}$ & $\,1.70$ &  & $3^{++}$ & $3.66$\\\hline
$2^{++}$ & $2.39$ &  & $1^{--}$ & $3.81$\\\hline
$0^{-+}$ & $2.55$ &  & $2^{--}$ & $4.0$\\\hline
$1^{-+}$ & $2.96$ &  & $3^{--}$ & $4.19$\\\hline
$2^{-+}$ & $3.04$ &  & $2^{+-}$ & $4.22$\\\hline
$3^{+-}$ & $3.60$ &  & $0^{+-}$ & $4.77$\\\hline
\end{tabular}

\end{center}

\section{From light to heavy glueballs}

First, we review the status of the three lightest glueballs of Table 1, for
which some candidates exists.

\textit{Scalar glueball: }The resonances $f_{0}(1500)$ and $f_{0}(1710)$ were
investigated as glueball's candidates in various works
\cite{weingarten,close,stani,chenlattice,rebhan}. In Ref. \cite{stani} the
glueball (as a dilaton) was studied within the so-called extended Linear Sigma
Model (eLSM) \cite{dick}. Quite remarkably, there is only an acceptable
scenario: $f_{0}(1710)$ is mostly gluonic. This is in agreement with the
original lattice work of Ref. \cite{weingarten}, with the recent lattice study
of $j/\psi\rightarrow\gamma G$ in Ref. \cite{chenlattice}, and also with the
AdS/QCD study in Ref. \cite{rebhan}. In conclusion, there is mounting evidence
that $f_{0}(1710)$ is predominantly the scalar glueball.$\ $

\textit{Tensor glueball: } In Ref. \cite{anisovich} it was shown that
$f_{J}(2220)$ does not lie on the Regge trajectories. Its mass fits well with
lattice (see Tab. 1), it is narrow, the $\pi\pi/KK$ ratio agrees with flavour
blindness \cite{tensor}, and no $\gamma\gamma$ decay was seen, hence it is a
good candidate to be the tensor glueball. Yet, the experimental assessment of
this resonance is necessary.

\textit{Pseudoscalar glueball: }The pseudoscalar glueball has been
investigated in a variety of scenarios, see the review \cite{masoni}. In some
works, e.g. Ref. \cite{tichy}, the pseudoscalar glueball was assigned to the
resonance $\eta(1405),$ but it is not clear if $\eta(1405)$ and $\eta(1475)$
are two independent states (see the recent discussion in Ref. \cite{exps} and
refs. therein). Moreover, the lattice mass is about $2.6$ GeV, i.e. 1 GeV
heavier. In Ref. \cite{psg} the decays of an hypothetical pseudoscalar
glueball (linked to the chiral anomaly \cite{salopsg}) with a mass of about
$2.6$ GeV were studied within the eLSM: the decay channels into $KK\pi$ and
$\eta\pi\pi$ are dominant. (For an independent AdS/QCD calculation, see
\cite{rebhanpsg}). A possible experimental candidate is the state $X(2370)$
measured by BES \cite{bespsg}, yet future measurements are needed.

For the other glueballs listed in\ Table 1, no candidate is presently known.
Very recently, two theoretical studies have been performed with the aim of
helping future experimental search.

\textit{Pseudotensor glueball:} in\ Ref. \cite{ptg} the decays of a
pseudotensor glueball, a putative resonance $\eta_{2}\equiv G(3040),$ has been
studied in a flavour-invariant hadronic model: sizable decay into $K_{2}%
^{\ast}(1430)K$ and $a_{2}(1320)\pi$ are predicted (they are enhanced by
isospin factors). Moreover, decays into a vector and a pseudoscalar mesons
vanish at leading order, hence $\Gamma_{G\rightarrow\rho\pi}=\Gamma
_{G\rightarrow K^{\ast}(892)K}=0$.

\textit{Vector glueball:} The vector glueball is interesting since it can be
directly formed in $e^{+}e^{-}$ scattering. Up to now, the search for
candidates was not successful \cite{besvg}. The decays of a vector glueball
(called $\mathcal{O}\equiv\mathcal{O}\mathbb{(}3810)$) using the eLSM have
been studied in\ Ref. \cite{vg} (for previous theoretical works, see Ref.
\cite{robson}). Three interaction terms have been considered. While the
intensity of the corresponding coupling constants cannot be determined, some
decay ratios are predicted. In\textbf{ }the first two interaction terms (which
are also dilatation invariant, then should dominate) the main decay modes are
$\mathcal{O}\rightarrow b_{1}\pi\rightarrow\omega\pi\pi$ (first term) as well
as $\mathcal{O}\rightarrow\omega\pi\pi$ and $\mathcal{O}\rightarrow\pi
KK^{\ast}(892)$ (second term). The third interaction terms, which breaks
dilatation invariance, predicts decays into vector-pseudoscalar pairs, in
particular $\mathcal{O}\rightarrow\rho\pi$ and $\mathcal{O}\rightarrow
KK^{\ast}(892)$.

Both the pseudotensor and the vector glueballs (as well as other heavy
glueballs) can be experimentally produced in formation processes at the future
PANDA experiment \cite{panda}.

\section{Glueballs' widths via large-$N_{c}$ considerations}

The widths of glueballs are presently unknown. The scalar and the pseudoscalar
glueball are somewhat special, since they are linked to the trace and axial
anomalies. However, for the other glueballs, a (rough!) estimate using
large-$N_{c}$ considerations may be helpful (for reviews on the large-$N_{c}$
approach see \cite{thooft}).

First, we recall that the dominant decay of a $\bar{q}q$ meson into two
$\bar{q}q$ states scales as $1/N_{c}.$ A typical example is that of the
$\rho(770)$ meson ($\rho(770)^{+}\equiv u\bar{d}$ decays trough creation of a
$\bar{u}u$ or $\bar{d}d$ pair from the vacuum that recombine in $\pi^{+}%
\pi^{0}$):%
\begin{equation}
\Gamma_{\rho\rightarrow\pi\pi}\propto\frac{1}{N_{c}}\text{ and }\Gamma
_{\rho\rightarrow\pi\pi}^{\exp}=147.8\pm0.9\text{ MeV}.
\end{equation}
Similarly, the vector kaonic state $K^{\ast}(892)$ decays into $K\pi$ is
regulated by a similar strength $\Gamma_{K^{\ast}(892)\rightarrow K\pi}^{\exp
}=50.3\pm0.8$ MeV (it is smaller only due to phase space). From the tensor
sector: $\Gamma_{f_{2}^{\prime}(1525)\rightarrow KK}=67.4\pm8.9$ MeV. Other
examples are in the PDG. Interestingly, when going to higher masses, the
\textit{qualitative} picture does not change. For instance, for the $\bar{c}c$
state $\psi(4040)$: $\Gamma_{\psi(4040)\rightarrow DD}^{\exp}=80\pm10$ MeV.
Summarizing, whenever a strong (OZI allowed) decay of a conventional $\bar
{q}q$ meson $M$ into two conventional mesons $M_{1}$ and $M_{2}$ is
kinematically allowed, one has:
\begin{equation}
\Gamma_{M\rightarrow M_{1}M_{2}}^{\text{OZI-allowed}}\propto\frac{1}{N_{c}%
}\text{ and }\Gamma_{M\rightarrow M_{1}M_{2}}^{\text{OZI-allowed}}%
\sim100\text{ MeV}.
\end{equation}
Clearly, there are modifications when threshold effects are important and/or
certain quantum numbers are considered, but the whole picture is rather stable.

Next, let us consider OZI suppressed decay \cite{ozi}, which occur through
annihilation of the original $\bar{q}q$ pair into gluons, which then reconvert
into $\bar{q}q$ mesons. In the large-$N_{c}$ language, the scaling
$\Gamma_{\bar{q}q\rightarrow MM}^{\text{OZI-suppressed}}\propto N_{c}^{-3}$
holds. A nice example is provided by the decay of the tensor state
$f_{2}^{\prime}(1525)$ into $\pi\pi$. This state is predominantly $\bar{s}s$,
hence this transition goes as $N_{c}^{-3}$.\ Experimentally:%
\begin{equation}
\Gamma_{f_{2}^{\prime}(1525)\rightarrow\pi\pi}\propto\frac{1}{N_{c}^{3}}\text{
and }\Gamma_{f_{2}^{\prime}(1525)\rightarrow\pi\pi}^{\exp}=0.62\pm0.14\text{
MeV}%
\end{equation}
which is a factor $100$ (!) smaller than $\Gamma_{f_{2}^{\prime}%
(1525)\rightarrow KK}$ (even if the phase space into $\pi\pi$ is large). A
very well known example in the heavy quark sector is given by the decay of the
$j/\psi$ meson into hadrons. The full hadronic decay (also suppressed by
$N_{c}^{-3}$) reads $\Gamma_{j/\psi\rightarrow\text{hadrons}}^{\exp}%
=0.081\pm0.002$ MeV. As shown in \cite{thooft}, the large-$N_{c}$ approach
naturally explains the validity of the OZI rule \cite{ozi} (and is actually
the only theoretical framework to derive it). Also in this case, various other
examples exist, such as $\Gamma_{\psi(2S)\rightarrow\text{hadrons}}^{\exp
}=0.280\pm0.07$ MeV and $\Gamma_{\chi_{c2}(1P)\rightarrow\text{hadrons}}%
^{\exp}=1.93\pm0.11$ MeV. In conclusion, one has the following estimate:%
\begin{equation}
\Gamma_{M\rightarrow M_{1}M_{2}}^{\text{OZI-suppressed}}\propto\frac{1}%
{N_{c}^{3}}\text{ and }\Gamma_{M\rightarrow M_{1}M_{2}}^{\text{OZI-suppressed}%
}\lesssim1\text{ MeV}.
\end{equation}
Let us now turns to a heavy glueballs above $2.6$ GeV. The large-$N_{c}$
scaling of a glueball's decay into two conventional mesons is given by
$\Gamma_{G\rightarrow MM}\propto N_{c}^{-2},$ thus one has the relation
\begin{equation}
1\text{ MeV }\sim\Gamma_{M\rightarrow M_{1}M_{2}}^{\text{OZI-suppressed}%
}\propto\frac{1}{N_{c}^{3}}<\Gamma_{G\rightarrow MM}\propto\frac{1}{N_{c}^{2}%
}<\frac{1}{N_{c}}\propto\Gamma_{M\rightarrow M_{1}M_{2}}^{\text{OZI-allowed}%
}\sim100\text{ MeV.}%
\end{equation}
Hence, we \textquotedblleft\textit{guess}\textquotedblright\ that the expected
width of a glueball lies between $1$ and $100$ MeV:%
\begin{equation}
\Gamma_{G\rightarrow MM}\sim10\text{ MeV.}%
\end{equation}
Such a width is relatively small to allow experimental detection of such
states (in particular at PANDA \cite{panda}). For the special case of the
vector glueball \cite{vg}, we expect that its width is in between that of
OZI-suppressed and OZI-allowed charm-anticharm decays of vector states:
$0.298$ MeV$=\Gamma_{\psi(2S)}^{\exp}<\Gamma_{\mathcal{O}\rightarrow
hadrons}<\Gamma_{\psi(4040)\rightarrow DD}^{\exp}=80$ MeV. Namely, the vector
glueball $\mathcal{O}$ is at an intermediate stage between three gluons and
quarks (in order $\mathcal{O}$ to decay, gluons have to completely annihilate
into quarks). Moreover, the fact that at least three valence gluons are
contained in the vector glueball, is a further hint that its decay should not
be large.

\section{Conclusions}

Glueballs are expected to exist but were not yet found in experiments. In this
work, we have briefly reviewed the status of some candidates and presented
predictions for the heavy pseudotensor ad vector glueballs. Moreover, we have
discussed the possible width of an heavy glueball, obtaining the heuristic,
rough estimate of about $10$ MeV. Experimental searches at low energies in the
experiments GlueX \cite{gluex} and CLAS12 \cite{clas12} at Jefferson Lab (see
also \cite{gutschenew}) and at high energy at the ongoing BESIII
\cite{bespsg,bes} and at the future PANDA \cite{panda} experiments are
expected to improve our understanding.

\bigskip

\textbf{Acknowledgments}: The author acknowledges support from the Polish
National Science Centre NCN through the OPUS project no. 2015/17/B/ST2/01625.

\end{document}